\documentclass[conference]{IEEEtran}
\IEEEoverridecommandlockouts
\usepackage{amsmath,amssymb,amsfonts}
\usepackage{algorithmic}
\usepackage{graphicx}
\usepackage{textcomp}
\usepackage{hyperref}
\usepackage[dvipsnames]{xcolor}
\def\BibTeX{{\rm B\kern-.05em{\sc i\kern-.025em b}\kern-.08em
    T\kern-.1667em\lower.7ex\hbox{E}\kern-.125emX}}
\usepackage{array}  
\usepackage{makecell}
\usepackage{enumitem}
\usepackage{tikz}
\usetikzlibrary{shapes.geometric}
\usetikzlibrary{arrows}

\usepackage[backend=biber,style=ieee]{biblatex} 
\begin{document}

\title{Quantum Computing Education in Latin America: Experiences and Strategies}

\author{\IEEEauthorblockN{Laura Tenjo-Patiño}
\IEEEauthorblockA{\textit{Departamento de Física y Química} \\
\textit{Universidad Nacional de Colombia}\\
Manizales, Colombia \\
ltenjo@unal.edu.co}
\and
\IEEEauthorblockN{Cristian E. Bello}
\IEEEauthorblockA{\textit{Departamento de Matemáticas} \\
\textit{Universidad Nacional de Colombia}\\
Medellín, Colombia \\
crbellor@unal.edu.co}
\and
\IEEEauthorblockN{Alcides Montoya C.}
\IEEEauthorblockA{\textit{Departamento de Física} \\
\textit{Universidad Nacional de Colombia}\\
Medellín, Colombia \\
amontoya@unal.edu.co}
}

\maketitle

\begin{abstract}
Quantum computing is a rapidly advancing field with the potential to drive scientific, educational, and technological development. However, it faces a significant shortage of qualified experts, creating an urgent demand for skilled professionals. In Latin America, quantum education remains in its early stages, further widening the regional talent and access gap due to limited educational infrastructure and financial constraints. This work presents an initiative to integrate quantum computing into higher education in Latin America through the application of the European Competence Framework for Quantum Technologies, which offers a standardized approach to defining competency requirements and assessing essential skills in the field. We propose introductory courses aligned with the framework’s guidelines, designed to meet industry standards and reach a broad audience. Additionally, we introduce supplementary resources, including team dynamics and evaluation methodologies, to enhance the educational ecosystem. These initiatives aim to create a sustainable and comprehensive quantum education model across Latin America, to close the regional skills gap, foster inclusivity, and prepare a diverse workforce to contribute innovative solutions on the global stage.
\end{abstract}
\section{Introduction}
Quantum computing (QC), an emerging technology that is rapidly advancing in the research field \cite{li2001quantum}, holds the promise to revolutionize society as a significant breakthrough in modern IT in the coming years \cite{rietsche2022quantum}, \cite{jenkins2022quantum}. Its importance is evident from the substantial investments by developed nations and leading companies worldwide in research and development, aimed at achieving practical quantum computers. This captures the attention of individuals, countries, and governments alike due to its promising applications in almost every area imaginable \cite{rietsche2022quantum}, \cite{kaku2023quantum}. 

This field is profoundly multidisciplinary, requiring expertise in various fields for its development. As quantum technologies (QT) become increasingly important in industry, there is a growing demand for a skilled quantum workforce \cite{greinert2024advancing}. However, there is a significant global shortage of qualified professionals in QC, creating an urgent need for a new cohort of engineers, scientists, and programmers who can leverage quantum computers to tackle real-world challenges and bring innovative problem-solving skills to the field \cite{amin2019needs}.

The barriers to entry in QC are quite high due to factors including the need for specialized knowledge and the scarcity of experts in the field \cite{rietsche2022quantum}. The current talent gap reflects the disparity between the growing number of job openings for quantum computing specialists and the relatively small number of graduates prepared to fill these roles each year \cite{masiowski2022quantum}. To address this quantum divide, the first step is to enhance QC education, as it can equip students with the necessary understanding to meet the increasing demand and bridge the talent gap. However, this must take into account the challenge that understanding QC is not very easy without a foundation in quantum mechanics and mathematics \cite{amin2019needs}. 

While the workforce development issue is global, Latin America lags behind due to deficiencies in its educational systems and lack of financial support, with knowledge primarily stemming from individual efforts. On one hand, QC is increasingly gaining recognition within Latin America, and the region is following the lead of other countries through various events such as conferences and seminars. Additionally, there are organizations and communities that provide a supportive environment and guidance on the topic. But on the other hand, the situation in higher education is less favorable. 

Although the community works to raise awareness and provide basic tools to build expertise, university-level courses are essential for individuals to be qualified and competitive on a global scale. While there are courses and master’s degrees at various universities that cover quantum-related subjects, they are neither common nor widely explored, focused specifically on QC, or accessible to a broader audience. One of the main challenges in establishing such courses is the lack of trained faculty members with expertise in quantum computing, making it difficult to integrate the topic into curricula. Furthermore, the lack of governmental support results in fewer companies and institutions providing a rigorous environment for research and practical application, making it challenging for students to apply theoretical knowledge in real-world scenarios.

According to the $2023$ Global Quantum Computing Map \cite{BBVA_Quantum}, the United States, China, and Europe are leading the development of QC, while Japan, South Korea, Singapore, and Australia are also active participants in the global race. Venegas-Gomez \cite{venegas2020quantum} presents a global effort map for public funding in QT, an effort that has been enhanced by the establishment of national programs worldwide. However, no Latin American country shows significant investments in QC that would warrant notable mention on the map, nor are they included in the global distribution of companies working in QT per country.

This economic context and the region’s underrepresentation in the field also underscore the need for a more specialized approach to QC education in Latin American universities. Such an approach would improve educational outcomes and stimulate innovation and company growth, contributing to a more competitive quantum technology sector.

\subsection{The European Competence Framework for Quantum Technologies}

\noindent The data underscores not only the urgency of preparing the future workforce, but also the need of tailored educational programs addressing emerging demands \cite{greinert2023future}. It is crucial to understand the specific qualification required by the industry for various job roles to create targeted courses, specialized tracks, and degree programs at universities which enhance students’ competitiveness for quantum-related positions \cite{hughes2022assessing}. To achieve this, the European Competence Framework for Quantum Technologies (CFQT) \cite{Greinert2024} was developed following a study aimed at collecting and identifying the competences, knowledge, and skills relevant for the emerging quantum workforce and its development \cite{greinert2024advancing}, \cite{greinert2023future}. This framework establishes standardized competency requirements for diverse roles in QT, ensuring transparency, scalability, and comparability in defining essential skills, and enabling consistent assessment across different courses or qualification \cite{greinert2023towards}, \cite{greinert2025extending}. Therefore, one viable solution for Latin American countries is to leverage this framework as a guide to ensure a comprehensive and standardized approach to QC education. 

To address this issue, a Latin American research group is committed to making quantum computing education open and accessible to students across the region. Their work focuses on integrating QC into local educational systems, with a strong emphasis on open science and inclusive learning. This is particularly relevant in Latin America, where educational resources in Spanish are limited or often paywalled, as most materials—such as courses, textbooks, programming tutorials, workshops, and games—are predominantly available in English \cite{maldonado2022quantum}, creating an additional barrier for non-English speakers. The group offers free courses via platforms like Twitch and YouTube and is planning to expand their course offerings in the near future. Their aim is to align these initiatives with the European Competence Framework for Quantum Technologies to ensure global relevance.

This study seeks to explore how the CFQT can be implemented to support the development of quantum computing education in Latin America, and what strategies may enhance its applicability in under-resourced educational contexts. 

\section{Methodology}

\begin{figure}[htbp!]
    \centering
    \includegraphics[scale=0.54]{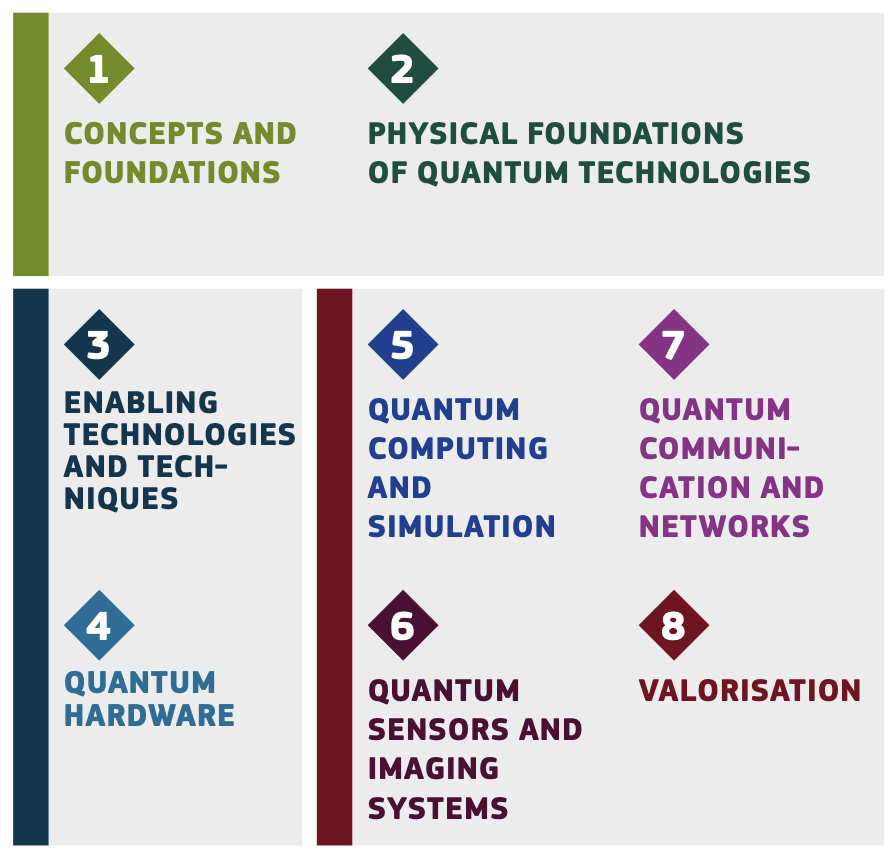}
    \caption{Framework domains overview with topics. Source: European Competence Framework for Quantum Technologies.}
    \label{fig:eightdomains}
\end{figure}

We will now present the implementation of the European Competence Framework for Quantum Technologies (version 2.5) to develop two initial courses. To begin, we will first explore its structure. The framework consists of three main parts. First, a content map (Fig.~\ref{fig:eightdomains}) with $8$ domains and $42$ subdomains, which provides a graphical overview of the structure and topics of QT.

The framework also contains proficiency levels from A1 to C2 (Tab.~\ref{tab:proficiency_levels}), analogously to the Common European Framework of Reference for Languages (CERF) \cite{CouncilofEurope2020} levels, to tailor education to specific target groups' needs and to determine job profiles. 

\begin{table}[htbp!]
\centering
\caption{Proficiency Levels}
\begin{tabular}{|c|c|}
\hline
\textbf{Proficiency Level} & \textbf{Keyword} \\
\hline
A1 & Awareness \\
A2 & Literacy \\
B1 & Utilisation \\
B2 & Investigation \\
C1 & Specialisation \\
C2 & Innovation \\
\hline
\end{tabular}
\label{tab:proficiency_levels}
\end{table}

\noindent The depth of knowledge required for each level are described, along with the time it takes to achieve it. These levels apply to three proficiency areas that encompass different competences, each being color-coded according to the domains it covers:

{
\renewcommand{\theenumi}{\Roman{enumi}}
\begin{enumerate}
    \item Quantum Concepts.
    \item QT Hardware and Software Engineering.
    \item QT Applications and Strategies.
\end{enumerate}
}

\begin{figure}[htbp!]
    \centering
    \includegraphics[scale=0.36]{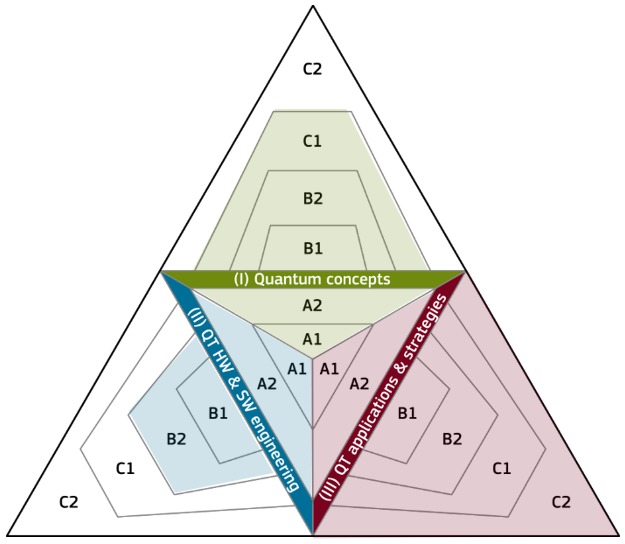}
    \caption{Framework's proficiency triangle with the three areas and the six proficiency levels.}
    \label{fig:prof_triangle}
\end{figure}

Using the proficiency levels and areas, the proficiency triangles (Part $2$) illustrate the general qualifications covered by a person or addressed in a course, effectively outlining its learning objectives. Fig.~\ref{fig:prof_triangle} illustrates an example of a proficiency triangle created using the Qualification Profile Creation Tool, provided by the CFQT authors, where the individual achieved C1, B2, and C2 proficiency levels in the three respective areas.

\noindent Finally, Part $3$ consists of qualification profiles (from P1 to P9) that provide an overview of common personal QT-specific qualifications relevant to industry. The framework shows additional insights on the profiles, including descriptions of the highest levels addressed, a general explanation of the profile, example personas and related recommendations.

We now introduce the designed courses, which include general information and descriptions. Subsequently, in alignment with the framework’s sections, parameters, and specifications, further aspects are detailed, such as the course content and the target qualification profile that students are expected to achieve. This adaptation involves the following elements:

\begin{enumerate}
    \item The proficiency triangle for the course's specific qualification profile, illustrating the proficiency level students are expected to reach in the three main areas by the end of each course.
    \item A general overview of the expected outcomes, outlining what students should be able to accomplish with the proficiency levels attained in each of the three areas covered by the framework.
    \item A selection of topics and subtopics from the $42$ available options, carefully chosen to align with the course objectives.
    \item Following the framework’s guidelines to combine proficiency levels with an appropriate selection and adaptation of content (sub)domains to define a course's objectives or an individual’s qualifications, we developed a detailed target profile. This profile outlines the specific knowledge and skills students will acquire in each area. The qualification profile was designed based on the framework's recommendations, while being tailored to fit the objectives of each course.
\end{enumerate}
 
 The section ``Extending the Framework and Additional Strategies for the Latin American Context" includes both adaptations based on the framework and complementary strategies tailored to the region’s needs. Specifically, building on the framework’s guidance for course development, we suggest supplementary elements —work teams and evaluation methodologies— to augment the courses, ensuring they meet the framework’s standards and enhance the learning experience, while also discussing broader strategies that enhance accessibility and strengthen quantum education in Latin America.

\definecolor{cOne}{RGB}{111,138,13}

\definecolor{cFive}{RGB}{19,64,148}

\definecolor{cEight}{RGB}{121,3,30}

\definecolor{NavyBlue}{RGB}{0, 48, 135}

\definecolor{BrickRed}{RGB}{203, 65, 84}

\definecolor{Cerulean}{RGB}{0, 123, 167}

\definecolor{YellowGreen}{RGB}{154, 205, 50}

\newcommand{\diamondnumber}[2]{%
    \tikz[baseline={(0, 0.1)}]{
        \draw[fill=#1, draw=none, text=white] 
        (0,0) -- (0.25,0.25) -- (0,0.5) -- (-0.25,0.25) -- cycle;
        \node[align=center, text=white, scale=0.8] at (0, 0.25) {#2};
    }%
}

\newcommand{\smallcirclednumber}[2]{%
    \tikz[baseline=(char.base)]{
        \node[shape=circle, fill=#1, text=white, draw, inner sep=1pt, scale=0.7] (char) {#2};
    }%
}

\newcommand{\rectanglenumber}[2]{\tikz[baseline=(char.base)]
{\node[shape=rectangle, fill=#1, text=white, draw, minimum width=1.8em, minimum height=1.3em, inner sep=0pt] (char) {#2};}}

\subsection{Proposed Courses and Implementation Strategy}

\subsubsection{Basic Summer/Winter Introductory Course}

\textbf{General Information:}
\noindent
\begin{itemize}
    \item Duration: Four weeks (two hours per week).
    \item Prerequisites: None.
    \item Description: This introductory course provides students with an initial exposure to the current state of QT. It offers a broad perspective on how different quantum technologies interconnect (not solely covering QC), and is suitable for non-STEM degrees and professionals from diverse fields working in quantum-related projects, providing basic knowledge to those without a technical background. It aims to familiarize students with tangible experiences through the use of games, visual activities, or simulations to help grasp key concepts, promoting conceptualization and intuitive understanding rather than focusing on extensive mathematical competence.
\end{itemize}
\vspace{4pt}

\begin{figure}[htbp!]
    \centering
        \includegraphics[scale=0.15]{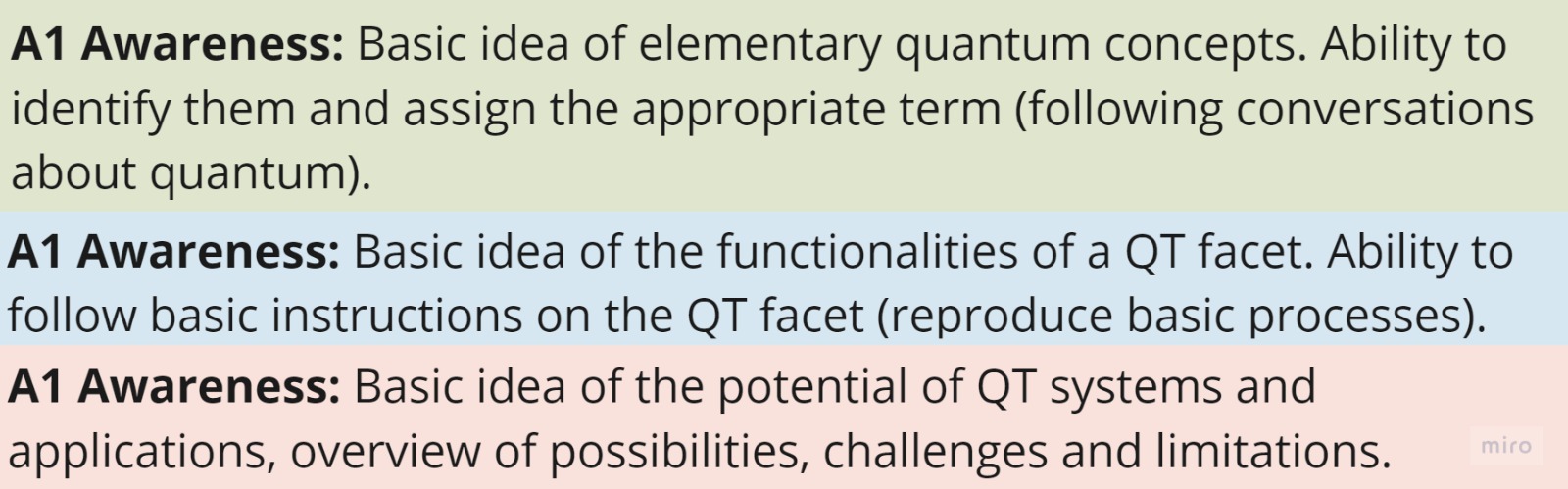}
        \caption{Target proficiency triangle for the first course and general descriptions of the target proficiency level for each area. These concise descriptions were adapted from the more detailed descriptions in the CFQT.}
        \label{fig:3}
\end{figure}

\begin{figure}[htbp!]
    \centering
    \includegraphics[scale=0.4]{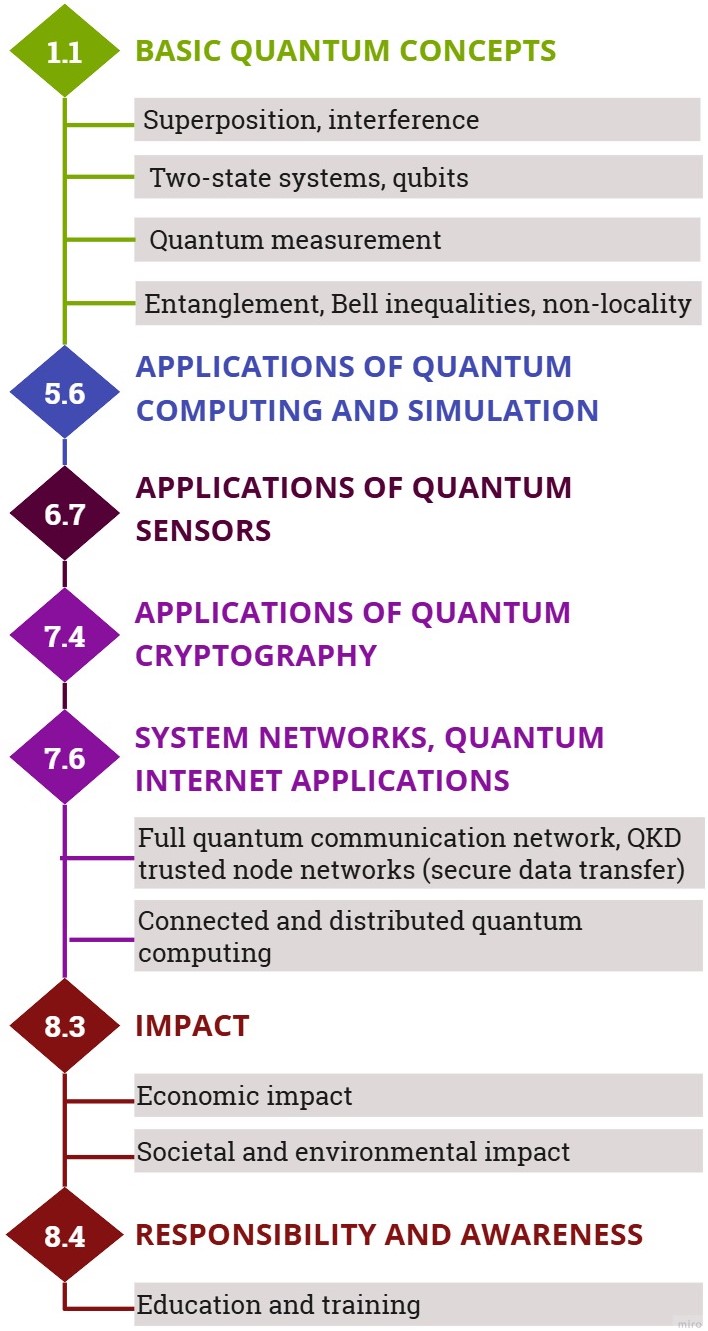}
    \caption{Subdomains and topics to be covered in the basic course. Adapted from the European Competence Framework for Quantum Technologies.}
    \label{fig:contents1}
\end{figure}

\noindent \textbf{Adaptation to the CFQT:}\\
For this course, the primary focus aligns with the \smallcirclednumber{NavyBlue}{P1} qualification profile (``QT Aware Person") and the Quantum Background domain \diamondnumber{cOne}{1} from the content map, which defines the target profile of a \textbf{\textcolor{blue}{General QT aware person}}. However, to provide a more comprehensive understanding, the course also incorporates selected subdomains related to Quantum Technology Applications (subdomains 5.6, 6.7, 7.4 and 7.6), offering participants an overview of the broader landscape of QT. Additionally, topics from domain 8 are included to highlight impact and awareness, completing a general overview. This approach ensures that learners gain foundational knowledge while appreciating the potential applications across various fields. The description and qualifications are shown in Table~\ref{tab:qualifications1}. Fig.~\ref{fig:3} illustrates the proficiency triangle and its levels description. Fig.~\ref{fig:contents1} shows the chosen focus contents for the course.

\vspace{4pt}

\begin{figure}[htbp!]
    \centering
        \includegraphics[scale=0.41]{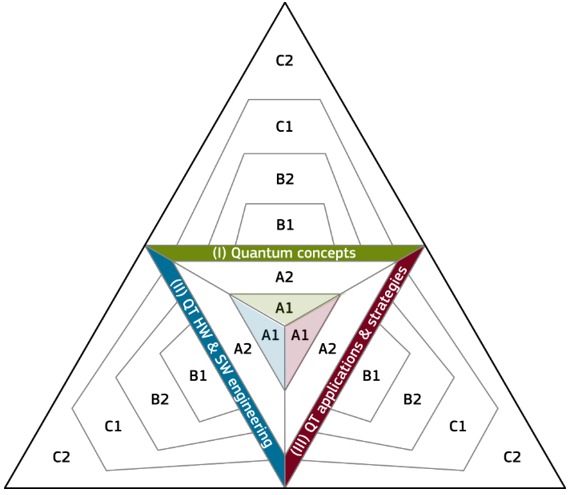}
\end{figure}
\begin{table}[htbp!]
\begin{center}
\caption{Persona's description and specific qualifications (First Course)}
\label{tab:qualifications1}
\begin{tabular}{ m{2cm} | m{5cm} } 
  \textbf{Target Qualification Profile} & General QT aware person: Has a general awareness of the current QT landscape, is familiar with key terms, has an overview of the possibilities, challenges and limitations of QT, and knows the basic quantum concepts in the context of QT functionalities.\\
  \\ 
  \hline
  \\ 
  \textbf{Specific Qualification} & Reproduce basic quantum terminology in the main QT areas (quantum communication, quantum sensors, quantum computing, quantum simulations). Reproduce basic functionalities of QT systems like quantum sensors, quantum simulators, control software. Recognize potential of QT, understanding its current limitations and future potential in solving complex problems, as well as its economic, societal, and environmental impacts.
\end{tabular}
\end{center}
\end{table}

\subsubsection{Undergraduate Elective Course}

\textbf{General Information:}
\noindent
\begin{itemize}
    \item Duration: Sixteen weeks ($4$ hours per week).
    \item Prerequisites: Basic Introductory Course and STEM basic courses such as Linear Algebra, Python Programming, Probability and Complex Variables.
    \item Description: This course shifts the focus more towards QC, offering deeper insights into its applications and challenges. The course combines both theoretical and practical aspects of QC, providing opportunities to interact with professionals and engage in problem-solving tasks. Despite the advanced content and mathematical formalism, the course is designed to be relatively accessible as a basic course. Combines content from the CFQT and a selected guidebook to ensure a QC-focused curriculum.
\end{itemize}
\vspace{4pt}

\begin{figure}[htbp!]
    \centering
        \includegraphics[scale=0.14]{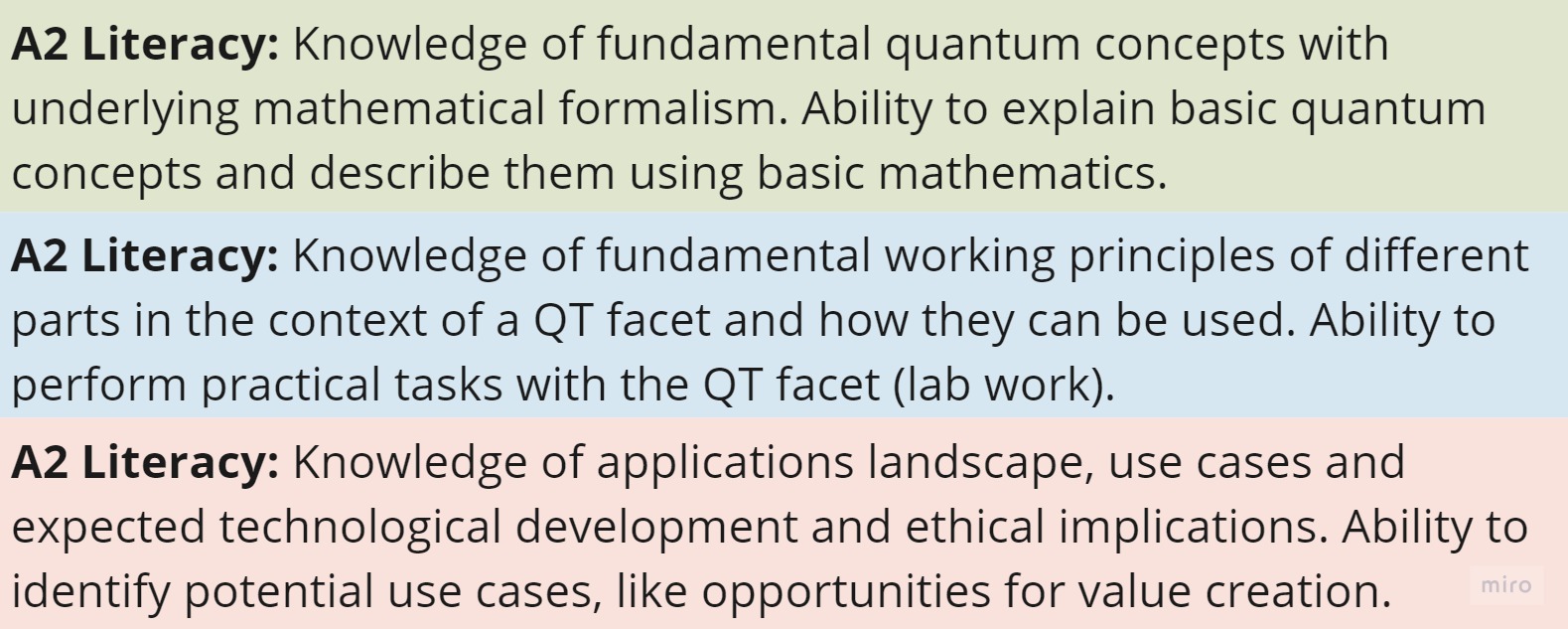}
        \caption{Target proficiency triangle for the second course and general descriptions of the target proficiency level for each area. These concise descriptions were adapted from the more detailed descriptions in the CFQT.}
        \label{fig:curso2}
\end{figure}

\begin{figure}[htbp!]
    \centering
    \includegraphics[scale=0.4]{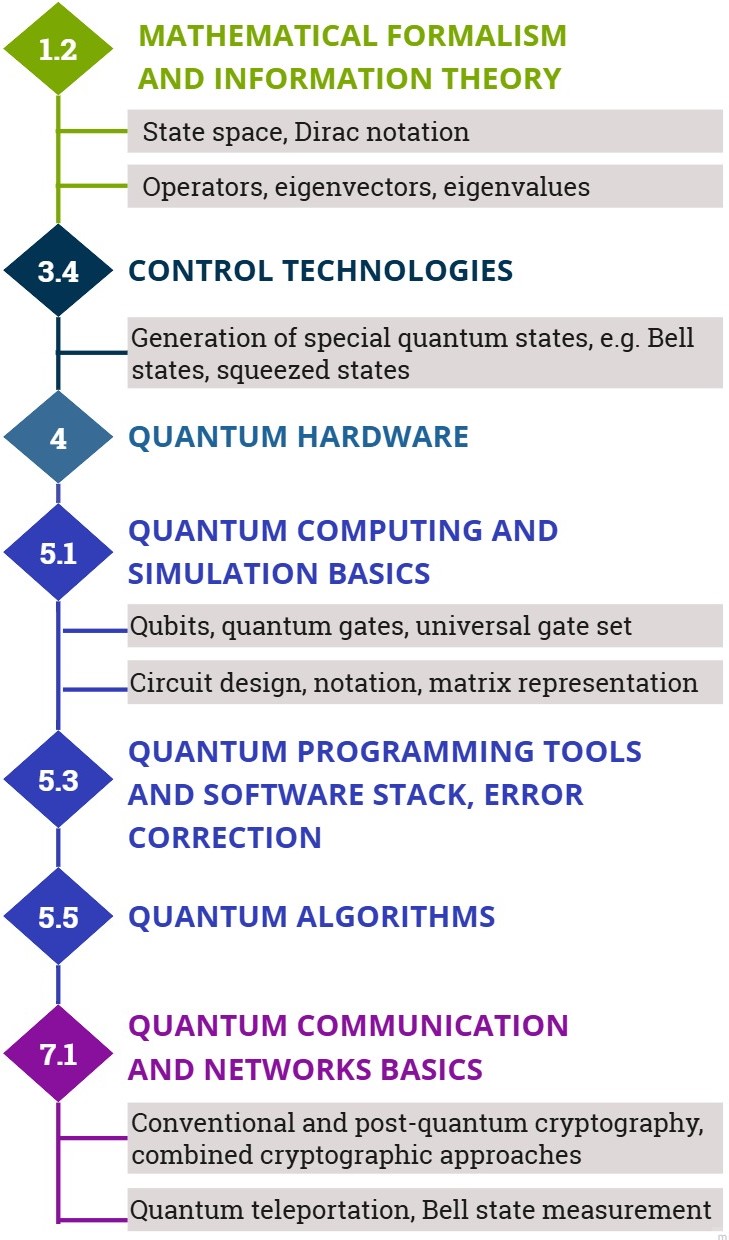}
    \caption{Subdomains and topics to be covered in the second course. Adapted from the European Competence Framework for Quantum Technologies.}
    \label{fig:contents2}
\end{figure}

\noindent\textbf{Adaptation to the CFQT:} 
\\
For this course, the combination of the \smallcirclednumber{NavyBlue}{P3} qualification profile (``QT Literate Person") with the Quantum Computing and Simulations domain \diamondnumber{cFive}{5} from the content map leads to the target profile of \textbf{\textcolor{blue}{QC literate person}}. The description and qualifications are shown in Table~\ref{tab:qualifications2}. Fig.~\ref{fig:curso2} illustrates the proficiency triangle and its levels description. Fig.~\ref{fig:contents2} shows the chosen focus contents.

\begin{figure}[htbp!]
    \centering
        \includegraphics[scale=0.4]{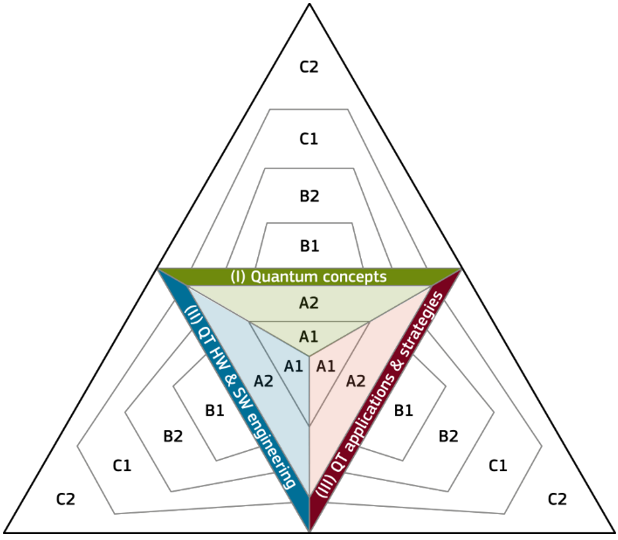}
\end{figure}
\begin{table}[htbp!]
\begin{center}
\label{tab:qualifications2}
\caption{Persona's description and specific qualifications (Second Course)}
\begin{tabular}{ m{2cm} | m{5cm} } 
  \textbf{Target Qualification Profile} & QC literate person: This person understands the language of quantum computing, possesses familiarity with key terminology, and can effectively communicate with experts and beginners in the field. They also have a foundational grasp of the potential applications and challenges associated with QC.\\
  \\ 
  \hline
  \\ 
  \textbf{Specific Qualification} & Describe fundamental quantum computing concepts. Perform basic tasks on a QC facet like quantum programming languages and quantum algorithms. Identify value of QC like solving complex computational problems more efficiently than classical computers and recognizing potential industry impacts and benefits in different fields.
\end{tabular}
\end{center}
\end{table}

\textbf{Note:} To ensure alignment with the framework's example personas, recommended learning paths, and suggested general conditions for the course's qualification profile (P1 and P3 in our case), students can consult the general qualification profile described in the framework. Although no specific content combination may be applied, this section covers general profiles, and the specific ones should also adhere to these guidelines.

\subsection{Extending the Framework and Additional Strategies for the Latin American Context}

\subsubsection{Work Teams}
Analogously to the qualification profiles and example personas given in Part $3$ of the CFQT, examples of real work teams are integrated into the courses to illustrate applicability of students' future work, providing them with real-life scenarios of quantum applications. This approach also offers a broader perspective on career opportunities within the quantum field, allowing students to explore potential job prospects and the personal impact of working in this sector. In Latin America, where the quantum industry is still emerging, it is beneficial to tailor these examples to regional needs and economic sectors where quantum technologies could have a transformative impact.
    
A starting point is to use the already existing example personas in the framework, built by combining qualification profiles with a selection from the content map to lead to titles that specify the focus, and later new profiles or professions outside the field can be included. The following work team example does not necessarily reflect a real-life situation, but it serves for illustrative purposes to demonstrate how the framework's concepts could be adapted to profile team members in sectors relevant to Latin America.\\

\noindent \textbf{Example: Finance}

\renewcommand{\theenumi}{\Roman{enumi}}
\begin{enumerate}
    \item Application's Description: This project focuses on leveraging quantum computing to analyze financial risks in emerging economies. Unlike stable financial markets, Latin American economies often face high inflation rates and currency fluctuations.
    \item Objective: Develop and implement quantum algorithms to improve risk assessment models for volatile economies and optimize currency exchange strategies in markets with fluctuating rates, while considering various financial constraints and market dynamics.
    \item Team Collaboration:
    \begin{itemize}
        \item \textit{Quantum Algorithm Computer Scientist:} \smallcirclednumber{NavyBlue}{P6} + \diamondnumber{cFive}{5.5}\\
        Designs and implements quantum algorithms that integrate macroeconomic indicators into quantum-enhanced predictive models, using appropriate quantum programming tools and software stacks. Works with the Financial Market Analyst and the Quantum Computing Specialist to ensure algorithmic feasibility and alignment with financial objectives. 
        \item \textit{Quantum Computing Specialist:} \smallcirclednumber{NavyBlue}{P7} + \diamondnumber{cFive}{5.3}, \diamondnumber{cFive}{5.6}\\
        Provides expertise in QC hardware and software. Adapts quantum models to available cloud platforms (IBM Quantum, AWS Braket) and advises on the selection of QC techniques to suit the requirements of financial tasks. Works with the Quantum Algorithm Computer Scientist to ensure alignment between algorithmic requirements and quantum hardware capabilities.
        \item \textit{Financial Market Analyst:} \smallcirclednumber{NavyBlue}{P5} + \diamondnumber{cEight}{8.1}\\
        Provides expertise in economic trends, inflation, and currency markets, translating financial objectives and constraints into optimization criteria, and defining the performance metrics for quantum algorithms. Collaborates with the Quantum Algorithm Computer Scientist and the Quantum Computing Specialist to provide insights into financial data and market behavior. 
        \item \textit{Business Strategist:} \smallcirclednumber{NavyBlue}{P8} + \diamondnumber{cEight}{8.2}, \diamondnumber{cEight}{8.3}\\
        Develops business strategies for commercialization of quantum solutions in the finance sector. Conducts market analysis, identifies potential partners or clients, and studies the feasibility of integrating quantum models into Latin American financial regulations. Collaborates with the Financial Market Analyst to assess market opportunities and define strategies. Also works with government regulators, banks, and Latin American fintech startups to explore real-world applications. 
    \end{itemize}

\item Workflow Overview

\begin{itemize}
    \item \textbf{Data Collection and Economic Analysis:} The Financial Market Analyst compiles and processes financial data from Latin American economies, including inflation rates, interest rates, and currency exchange trends. This data is used to define risk assessment parameters and optimization constraints for quantum models.
    \item \textbf{Quantum Model Development:} The Quantum Algorithm Computer Scientist designs and implements quantum algorithms tailored to financial risk modeling. The Quantum Computing Specialist assists in selecting the most appropriate quantum algorithms and ensures that the models are computationally feasible.
    \item \textbf{Implementation and Testing on Cloud-Based Quantum Platforms:} The Quantum Computing Specialist adapts the models for execution on cloud-based quantum platforms, optimizing for hardware constraints. The models undergo an iterative testing phase, in which the outputs are validated against historical financial data. The Financial Market Analyst assesses the accuracy of the quantum predictions and provides feedback for model refinement.
    \item \textbf{Market Integration and Strategic Deployment:} The Business Strategist evaluates the potential for commercial adoption of the quantum models in financial institutions. The final models are integrated into financial decision-making platforms, ensuring that they provide tangible value to financial institutions operating in emerging markets.
\end{itemize}

\end{enumerate}

\vspace{5pt}
   
\subsubsection{Evaluation}

Building on the framework’s guidance for course development, we suggest an evalaution scheme, ensuring it meets the framework’s standards and enhances the learning experience. Based on the proficiency level descriptions for the three areas - Quantum Concepts, QT HW and SW Engineering, and QT Applications and Strategies - which detail the expected knowledge and skills, course evaluation tasks can be developed to assess student learning. By integrating these tasks, clear metrics and assessment tools can be established to measure both the knowledge (K) and skills (S) outlined in the framework.\\

\noindent \textbf{Example: Basic Summer/Winter Introductory Course Evaluation}

\renewcommand{\theenumi}{\Roman{enumi}}
\begin{enumerate}

    \item Quantum Concepts (Target level: \rectanglenumber{YellowGreen}{A1} - Awareness)\\
    
    \noindent \textit{Objective:} 
    \begin{itemize}
        \item Evaluate students' ability to reproduce basic quantum concepts and terminology.
    \end{itemize}
    \noindent \textit{Knowledge Assessment (K):} Multiple-choice exam.
    \begin{itemize}
        \item Questions that assess the understanding of basic concepts (qubits, superposition, quantum measurement) and quantum terminology.
        \item Example: ``What is the concept of quantum superposition? (a) The ability of a particle to exist in a combination of different states simultaneously. (b) The ability of a photon to interact in such a way that its quantum state cannot be described independently.”
    \end{itemize}
    \noindent \textit{Skills Assessment (S):} Concept identification tasks.
    \begin{itemize}
        \item Students must read a popular science article or watch a video related to QT and then identify and explain the quantum terms discussed.
        \item Example: After reading an article, students highlight each quantum concept they recognize (e.g., entanglement, wave function collapse) and write a brief description of each one.
    \end{itemize}

    \item QT Hardware and Software Engineering (Target level: \rectanglenumber{Cerulean}{A1} - Awareness)\\

    \noindent \textit{Objective:}
    \begin{itemize}
        \item Evaluate the basic understanding of the functionalities of quantum HW/SW components, and the ability to reproduce basic functionalities of a QT facet and follow basic instructions.
    \end{itemize}
    \noindent \textit{Knowledge Assessment (K):} Workshop.
    \begin{itemize}
        \item Questions about the basic operation of hardware components (e.g., qubits, quantum gates) and software (basic quantum algorithms).
        \item Example: ``What is a qubit, and how does it differ from a classical bit? (a)  A qubit can exist in a superposition of both 0 and 1 states. (b) A qubit follows classical computing rules."
    \end{itemize}
    \noindent \textit{Skills Assessment (S):} Guided practice.
    \begin{itemize}
        \item Students follow a series of instructions to simulate a basic quantum circuit using a software simulator (e.g., Qiskit).
        \item Example: Guided hands-on programming practice - Circuits, application of quantum gates (Hadamard, Pauli gates).
    \end{itemize}

    \item QT Applications and Strategies  (Target level: \rectanglenumber{BrickRed}{A1} - Awareness)\\

    \noindent \textit{Objective:} 
    \begin{itemize}
        \item Evaluate students' ability to recognize the potential of QT applications and critically engage in public discussions about these technologies.
    \end{itemize}
    \noindent \textit{Knowledge Assessment (K):} Short essay.
    \begin{itemize}
        \item Students write a 500-word essay on a specific quantum application (e.g., quantum cryptography, quantum sensors) and discuss its possibilities, challenges, and limitations.
        \item Example: ``Discuss the potential applications of quantum computing in drug development."
    \end{itemize}
    \noindent \textit{Skills Assessment (S):} Discussion forum.
    \begin{itemize}
        \item Participation in an online forum where recent news or developments in QT are discussed. The evaluation focuses on their ability to follow and engage in conversations about these applications with a basic but critical understanding.
        \item Example: ``Debate the potential use of quantum computing in cybersecurity and whether you think media claims are exaggerated."
    \end{itemize}

    \item Overall Evaluation Proposal: The aim is to ensure a balanced evaluation of both theoretical understanding and practical application, in alignment with the framework's competencies. Slightly more weight is given to the first area, as the goal is to ensure students have a strong theoretical foundation in quantum concepts. This will provide them with the necessary background to better focus on software, hardware, and applications in subsequent courses. Therefore, in each of the three areas, a greater emphasis has been placed on knowledge assessment rather than skills assessment. This is illustrated in Tab.~\ref{tab:evaluation}.

\begin{table}[htpb!]
    \centering
    \caption{Evaluation Criteria and Their Percentages}
    \begin{tabular}{|l|c|}
        \hline
        \textbf{Assessment} & \textbf{Percentage} \\
        \hline
        Multiple-choice exam & 25\% \\
        Concept identification task & 15\% \\
        Workshop & 20\% \\
        Guided practice & 10\% \\
        Short essay & 20\% \\
        Discussion forum participation & 10\% \\
        \hline
    \end{tabular}
    \label{tab:evaluation}
\end{table}

\end{enumerate}
    
\subsubsection{Cloud-based Infrastructure as a Core Component}

Given the limited access to real quantum hardware in Latin America, the courses are designed around cloud-based quantum platforms such as IBM Quantum and Amazon Braket, as well as simulation tools like Qiskit and Cirq. This approach ensures that students not only gain experience with real quantum computing resources but also develop a strong foundational understanding through simulations. 

By integrating practical exercises and hands-on programming of quantum algorithms, the courses bridge the gap between theory and practice, allowing students to experiment with quantum circuits before working with actual hardware. Additionally, by familiarizing students with industry-standard tools and workflows, the courses enhance their preparedness for careers, whether in research or industry.

\subsubsection{PhD and Industry Guest Speakers}\label{subsubsec2}

The courses intend to integrate sessions with national and international PhD researchers and industry professionals, allowing students to gain insights into real-world challenges and how they are addressed. These sessions will also expose students to global career paths and collaboration opportunities, which is particularly valuable for Latin American students who often have limited direct exposure to the quantum industry. Speakers will discuss not only the personal impact of working in QC and the skills required to thrive in this sector but also share their experiences through interactive discussions, Q\&A sessions, and mentorship opportunities. 

By fostering a deeper connection between academia, industry, and students, this approach helps bridge the gap between Latin America and the global quantum ecosystem, enhancing career prospects and research opportunities in the region.

\subsubsection{Regional Collaboration}

As the course proposals are initially designed to be offered online, they have the potential to reach students from various universities, cities, and even countries across Latin America. This broader accessibility is particularly relevant in a region where quantum computing education is still developing and often concentrated in a few institutions. Expanding the scope and impact of these educational initiatives requires close collaboration between universities, research centers, and industry partners throughout Latin America. 

By sharing resources, expertise, and best practices, institutions can create a more comprehensive educational experience, ensuring that students have access to high-quality quantum education, regardless of their location.

To address the language barrier, developing and translating open-access materials, as well as creating Spanish-language learning resources, could serve as a starting point to enhance inclusivity and engagement in the field. Given that Spanish is widely spoken in Latin America, these efforts would benefit a significant portion of the population, while also laying the groundwork for future initiatives that consider the region’s linguistic diversity. 

Furthermore, fostering regional collaboration can help mitigate existing disparities in infrastructure and faculty training. Initiatives such as joint research projects, shared teaching materials, and inter-institutional workshops can enhance the learning experience and build a stronger quantum community within the region. 

\subsubsection{Global Partnerships}

Engaging with global quantum research communities, participating in joint projects, and facilitating student and faculty exchanges will enhance the quality of education and research, providing essential opportunities for knowledge exchange, and access to cutting-edge developments and resources, while ensuring that Latin American institutions remain at the forefront of quantum advancements. The courses’ funding could initially be requested from universities, but seeking support from existing international organizations that promote the dissemination of QC is also an objective to ensure sustainability and broader impact. 

\subsection{Outlook}

As a starting point, the first proposed course is under consideration for future testing, potentially being offered online by the UNAL’s research group during a semester break. This initial phase could also involve updates to their existing Quantum Computing course available on YouTube, aligning it with clearly defined learning objectives to meet the competency requirements outlined in the framework, ensuring that the course provides a structured pathway for students to progressively develop the necessary skills. A similar approach could be valuable for other universities looking to implement the framework, as feedback from students and faculty during this phase would be crucial for refining content, delivery methods, and evaluating the effectiveness of integrating quantum education into the actual curriculum in the future, offering online and onsite courses that can be taken with the free elective credits available to students, to implement them across a broader range of careers.

Developing further undergraduate and graduate specialization courses to gain deeper understanding and practical skills in specific areas is also an objective. The courses would focus on achieving the B proficiency level and qualification profiles from P4 to P6 depending on students’ interests; as reaching the C level and the P7, P8 and P9 profiles typically requires extensive work experience, such as a PhD or post-doctoral research. 

As QT continues to evolve rapidly, it is imperative to establish continuous professional development programs for educators, providing ongoing training and resources to keep instructors abreast of the latest advancements. However, in Latin America, where formal quantum education is still in its infancy, many universities lack faculty with prior expertise in QC. A potential future step in expanding QC education in the region is the development of a train-the-trainer program aimed at equipping educators with the necessary skills to teach quantum computing effectively. In addition, training courses based on the CFQT can be offered to students with more advanced knowledge, enabling them to become educators themselves. This approach not only helps increase the number of trained faculty but also facilitates the dissemination of knowledge to a broader student base, thereby expanding the reach and impact of QC education.

\section{Discussion}\label{sec3}

In addressing the challenge of quantum education and training, we aimed to close the gap in Latin America and the rest of the world with this first step of course creation, contributing to the unification of scientific and technological progress. We highlighted the contributions of the UNAL’s research group, an initiative that can serve as a model for other groups aiming to contribute to the field and its expansion. We also provided a practical and illustrative example of how to implement the European Competence Framework for Quantum Technologies, which not only guides the profiles and proficiency levels needed but also offers valuable information on needs, suggestions, and learning paths for each profile, useful to students once they end the course.

Additionally, we introduced the elements of ``Work Teams" and ``Evaluation" based on the framework to enhance the courses' development, important to illustrate applicability and unify assessment methods. These additions were made possible thanks to the framework’s clarity and comprehensibility, as it offers numerous tools that allow educators to maximize its potential and create additional elements for education. Further advancement of these tools will involve examining and creating additional examples and methodologies. 

A critical aspect that emerged during this process was the consideration of whether to introduce QC concepts through elective courses or integrate them into mandatory foundational STEM courses. While elective courses offer flexibility and cater to specific interests, integrating quantum content into core courses ensures a broader reach across disciplines, which could enhance interdisciplinary learning and prepare a wider range of students for the quantum era. However, we need to take into account the short term feasibility of this approach, given that universities tend not to be very flexible regarding the modification of existing courses.

Another consideration during the course creation process was whether to offer the courses to a broad range of majors, including non-STEM students where quantum technologies are becoming increasingly relevant, or to limit them to STEM fields. We decided to make the first introductory course accessible to all students, regardless of their academic background. However, the second course, which focuses on more advanced quantum physics and mathematics, does have prerequisites that non-STEM students typically do not fulfill. As we move forward, it will be important to assess whether this approach is effective for engaging a diverse student population (for example by comparing the academic performance of STEM and non-STEM students, which would reveal potential gaps in understanding), as addressing this challenge requires innovative teaching methodologies to make quantum concepts accessible and relevant to diverse majors. 

Alternatively, we could explore a different strategy where STEM students interested in specific applications (e.g., cancer research) learn relevant concepts from those fields to better communicate with professionals in interdisciplinary teams. For example, QC students could gain a foundational understanding of medical concepts to facilitate collaboration with doctors, rather than expecting non-STEM professionals to master quantum concepts.

\section{Conclusion}\label{sec4}

This work highlights the importance of tailoring quantum education initiatives to the specific needs of Latin America while leveraging established frameworks such as the CFQT. The introduction of tailored courses has the potential to significantly impact QC education across Latin America by promoting regional competitiveness and standardization. Our approach facilitates the creation of comprehensive courses that address immediate educational needs and ensure transparent learning objectives from the outset, providing students with a clear understanding of their learning goals. Through the integration of work teams and evaluation methodologies, we demonstrate how the framework can be effectively applied to course design and assessment in the region. Additionally, we recognize that addressing the educational challenges in Latin America requires complementary strategies beyond the framework itself. By incorporating these elements, we aim to create an inclusive quantum education model that enhances accessibility and prepares students for both local and international opportunities in the field. These efforts not only contribute to strengthening quantum education in Latin America but also offer insights that may be valuable for other regions facing similar challenges.

To ensure the long-term success and adaptability of these courses, collective experience and ongoing collaboration among educators, industry professionals, and academic institutions will be essential. Such a collaborative effort will enable continuous feedback and modifications, allowing QC education to evolve in ways that meet the diverse needs of students globally. In this context, future research should explore more innovative teaching methodologies and curriculum designs, as well as effective strategies for integrating theoretical concepts with hands-on applications, particularly in interdisciplinary fields.

One of the key challenges addressed is accessibility, both in terms of geographic reach and inclusivity for students lacking enough foundations in the field. By enhancing access to QC education across Latin America, we can leverage its diverse talent pool, fostering a new generation of professionals who bring varied perspectives and innovative solutions to the global stage. While the courses represent a step forward, further efforts are needed to ensure their scalability and adaptability. Potential limitations include relevant content selection and logistical challenges of delivering high-quality, practical QC education through online platforms.

In essence, by continuing to innovate in quantum education, expanding regional and global partnerships, and focusing on reducing barriers to entry, we can ensure that the next generation of professionals is well-equipped to meet the challenges and opportunities of the quantum future.

\section*{Acknowledgments}

We extend our gratitude to Franziska Greinert for her feedback and insights in creating the courses. We also thank her and her team for developing the framework, which provided a comprehensive foundation for our courses design and research direction. C.E.-B. and A.M.-C. acknowledge funding from the project “Centro de Excelencia en Computación Cuántica e Inteligencia Artificial”, HERMES code $52893$, UNAL.
\vspace{3pt}

\printbibliography

\end{document}